\documentclass[12pt, letterpaper, oneside]{article}

\usepackage[body={6.5in, 9in}, right=1in, top=1in]{geometry}
\usepackage{amsmath}
\usepackage{graphicx}

\newcommand\bcdot{\ensuremath{%
  \mathchoice%
   {\mskip\thinmuskip\lower0.2ex\hbox{\scalebox{1.5}{$\cdot$}}\mskip\thinmuskip}}%
   {\mskip\thinmuskip\lower0.2ex\hbox{\scalebox{1.5}{$\cdot$}}\mskip\thinmuskip}%
   {\lower0.3ex\hbox{\scalebox{1.2}{$\cdot$}}}%
   {\lower0.3ex\hbox{\scalebox{1.2}{$\cdot$}}}%
}

\begin{document}

\title{Effective dark matter fluid with higher derivative corrections. }
\author{N.A. Koshelev  \thanks{koshna71@inbox.ru}\\
{\small \it Ulyanovsk State Pedagogical University,} \\
{\small \it100 years V.I.Lenin's Birthday Sq.,4, Ulyanovsk 432700,
Russia} }

\maketitle

\begin{abstract}

The effective field theory for hydrodynamics allows to write the
action functional for fluid. In this paper, some simplest possible
higher derivative terms in the fluid action and the cosmological
consequences of their presence are considered. Particular
attention is given to dark matter, modelled as a dust with higher
derivative corrections. We study the conditions of absence of
singularities in the solutions of the background and perturbed
equations and investigate the evolution of perturbations in two
simple models of matter dominated Universe. There is a range of
parameters describing the higher derivative terms, in which the
short-wavelength perturbations of dark matter are suppressed and
the dark matter can be seen as fairly homogeneous on a
sufficiently small scale.

\end{abstract}

\section{Introduction.}
\label{sec:1}

Relativistic fluids play an important role in many areas of modern
cosmology  and astrophysics. They are widely used for the
description of various kinds of cosmological species. As an
example, the standard Big Bang cosmological model ($\Lambda$CDM)
contains non-clustering dark energy in the form of a small
positive cosmological constant $ \Lambda $ together with cold dark
matter considered as a perfect fluid with zero pressure. The
standard model fits well to the supernovae data  \cite{SN1a},
Baryon Acoustic Oscillations surveys \cite{BAO} and measurements
of the cosmic microwave background \cite{Hinshaw,Ade}. More
complicated models are also developed to avoid such theoretical
problems of $\Lambda$CDM cosmology as fine-tuning and
coincidences problems. These are primarily the models with dynamic
dark energy (see \cite{dynamic} for a review) and cosmological
models on the basis of modified theories of gravity (a recent
review can be found in \cite{modified}).  Generalized models of
dark matter alone are also of interest.

There are several different approaches to cosmological perfect
fluids. A promising way of describing a perfect relativistic fluid
is outlined by the pull-back formalism \cite{Andersson_Comer1} and
the effective field theory for hydrodynamics \cite{10116396}. In
these closely related techniques, a  barotropic perfect fluid can
be described by a set of three non-canonical scalar fields
satisfying certain symmetry conditions \cite{Dubovsky1}. The
advantage of these formalisms is the ability to write a perfect
fluid action in simple and convenient form that offers great
possibilities for generalizations. For example, the effective
field theory allows to describe a charged fluid \cite{Dubovsky2}.
Currently, both approaches are evolving to account for a range of
dissipative effects \cite{dissipative_fluid1, dissipative_fluid2}.

Cosmological applications of the variational principle for fluids
(and solids) were first considered in
\cite{ExtendedLCDM,12090464,ENW,BB}. In \cite{12090464} the
corresponding formalism  has been used  to constrain possible
deviations from Lorentz invariance in dark matter. It was applied
to propose and investigate new classes of inflationary models
\cite{ENW} as well as new form of possible coupling between dark
matter and quintessence \cite{PSC}.  The cosmological effective
field theory of multi-component fluids has been considered in
\cite{BBM}. The Brown's formulation of the variational principle
for relativistic fluids (with Lagrangian multipliers) \cite{Brown}
currently is intensively employed in new multi-fluid
\cite{150206613} and Scalar-fluid
theories\cite{150106540,150204030,15050755}.

The aim of this paper is to investigate some cosmological effects
of higher derivative corrections to the  action of perfect fluid
within the framework of the effective field theory. A sample of
such corrections has been considered in the Minkowski background
in \cite{Dubovsky2}. In a cosmological context, the effective
theory of fluids at next-to-leading order in derivatives and
implications for dark energy have been explored in \cite{NLO}.
This paper is organized as follows. In Section \ref{sec:2}, we
review the formalism used in the analysis and present our model.
The background solutions are investigated in Section \ref{sec:3}.
The Section \ref{sec:4} contains   the perturbed  Einstein
equations for scalar linear perturbations about spatially flat
FLRW background. We examine also the conditions of the absence of
instabilities in  radiation-dominated and  matter-dominated eras.
In Section \ref{sec:5}, the evolution of perturbations is studied
in the toy model and, by numerical methods, in a more realistic
model with baryonic matter. We conclude in Section \ref{sec:6}.

\section{Setup}
\label{sec:2}

Let us briefly review some basic concepts and
results of effective field theory for hydrodynamics. Initially,
this formalism has been developed as the theory of phonons in a
continuous medium \cite{Leutwyler1,Leutwyler2,Son,Dubovsky1}. It
has been applied also to rederiving, in a field theory language,
some findings of the pull-back formalism
\cite{KSG,Comer_Langlois}, which is based on earlier formulations
of the variational principle for fluids \cite{Taub,Carter}. Here
we mainly follow the papers \cite{Dubovsky1,Dubovsky2,BB}.

In continuum mechanics, the central role plays the concept of
fluid element. A fluid element is always situated at the same
fluid point moving with the velocity of the flow \cite{Pert}. The
fluid points can be marked with three labels $\varphi^a$
($a=1,2,3$), which are usually chosen to be a fluid point
coordinates at some initial time. At any time $t$ the coordinates
of the fluid element are given by $x^i=x^i(t, \varphi^a)$. This
equation thus defines the trajectory of the fluid element and
corresponds to a Lagrangian description of the fluid motion.

The inverse mapping
\begin{equation}
\label{comovmap}x^i\rightarrow \varphi^a(t,\mathbf{x})
\end{equation}
gives the labels of the fluid point that is located at spatial
point $\mathbf{x}$ at the moment $t$. In other words, the mapping
(\ref{comovmap}) provides an Eulerian  representation of fluid
flow. Once the labels are being attached to the fluid points, they
do not change under space-time coordinate transformations. Hence,
the functions $\varphi^a(t,\mathbf{x})$ are  scalar fields from
the observer's point of view.

The three-dimensional space of fluids points and the spatial space
are isomorphic, that allows to consider the map (\ref{comovmap})
as the transformation to a coordinate system in which the fluid is
at rest. Therefore, the fields $\varphi^i(t,\mathbf{x})$  can be
treated as the comoving coordinates of the fluid. This  leads to
the orthogonality condition ($u^\mu=dx^\mu/d\eta$, where $\eta $
is the proper time)
\begin{equation}
\label{ort}u^\mu\partial_\mu\varphi^a=0.
\end{equation}

The action of any physical theory which uses fields
$\varphi^a(t,\mathbf{x})$ should not depend on the ambiguity of
its definition. In particular, the action of the system should not
be changed within internal shifts and rotations:
\begin{eqnarray}
\label{symmetry1}\varphi^a & \rightarrow & \varphi^a+c^a,\\
\label{symmetry2}\varphi^a  & \rightarrow & O^a_{~b}\varphi^b,
~~~~~~O\in SO(3),
\end{eqnarray}
where $c^a$ and the matrix entries $O^a_{~b}$ are constant.

In addition, to distinguish a fluid from an isotropic solid, we
require the invariance under  volume preserving transformations
\begin{equation}
\label{symmetry3}\varphi^a\rightarrow f^a(\varphi), ~~~~~ \det
\left(\frac{\partial f^a}{\partial \varphi^b}\right)=1.
\end{equation}

\subsection{ Barotropic perfect fluid.}

The invariance under internal translations (\ref{symmetry1})
implies that the Lagrangian is a function of the quantities
$\varphi^a_{,\mu}\equiv\partial_\mu \varphi^a$ and their partial
derivatives. At lowest order in the derivative expansion, it will
involve only one derivative acting on each field $\varphi^a$. Then
the Lorentz invariance leads to the fact that the lowest order
Lagrangian should be built from the set of scalar quantities
\begin{equation}
\label{matrix}B^{ab}\equiv B_{ab} =g^{\mu\nu} \varphi^a_{,\mu}
\varphi^b_{,\nu} ,
\end{equation}
that form a matrix $B$.

There is only one independent function of entries of any $3 \times
3$ matrix, namely its determinant, which is invariant under the
internal transformations (\ref{symmetry2}), (\ref{symmetry3})
\cite{ENW}. Hence, the low-order action consistent with the
imposed symmetries is
\begin{equation}
\label{action_bf} S_{(bf)}=\int F(b)\sqrt{-g}d^4x ,
\end{equation}
where $F$ is a generic function and
\begin{equation}
\label{def_b}b = \sqrt{\det B}.
\end{equation}
From the viewpoint of field theory, this action is interpreted as
a low-energy one \cite{10116396}.

The orthogonality conditions (\ref{ort})  can be resolved to give
\cite{Dubovsky1,Dubovsky2}
\begin{equation}
\label{velocity}u^\mu = \frac{1}{b\sqrt{-g}}\epsilon^{\mu\alpha
\beta\gamma} \varphi^1_{,\alpha} \varphi^2_{,\beta} \varphi ^3
_{,\gamma},
\end{equation}
where the Levi-Civita symbol $\epsilon$ is defined by
$\epsilon^{0123}=+1$, and we use the usual normalization
\begin{equation}
\label{norm}u^\mu u_\mu=-1 .
\end{equation}
It is worth noting that the four-velocity  $u^\mu $ is invariant
to an arbitrary internal diffeomorphism \cite{10116396}.

We define the energy-momentum tensor as
\begin{equation}
T_{\mu\nu}=-\frac{2}{\sqrt{-g}}\frac{\delta S}{\delta g^{\mu\nu}},
\end{equation}
that gives for the low-order action (\ref{action_bf})
\begin{equation}
\label{bftensor}T_{\mu\nu (bf)} =Fg_{\mu\nu} -F_{,b}b
B^{-1}_{~~ab} \varphi^a_{,\mu} \varphi^b_{,\nu}.
\end{equation}

In view of equations (\ref{ort}) and  (\ref{velocity}), this is
the energy-momentum tensor of a perfect fluid
\begin{equation}
T_{\mu\nu}= (\rho+p)u_\mu u_\nu +p g_{\mu\nu},
\end{equation}
whose energy density and pressure are given by \cite{Dubovsky1}
\begin{equation}
\label{p_rho}\rho = -F ,~~~~~
p = F - F_{,b}b .
\end{equation}

Hence, the action (\ref{action_bf}) describes a barotropic perfect
fluid with the equation of state
\begin{equation}
p=w\rho,
\end{equation}
where
\begin{equation}
w\equiv\frac{p}{\rho} =\frac{F_{,b}}{F}b -1.
\end{equation}

The dust has zero pressure  $p = 0$, that leads to $F \propto b$.
The radiation fluid is described by $F\propto b^{4/3} $.
Generally, the power law dependence of the function $F(b)$
corresponds to the case of  barotropic fluid with a constant
parameter of state $w$.

Within  the reviewed above formalism there is conserved current
\begin{equation}
\label{current} J^\mu =b u^\mu .
\end{equation}
This vector field satisfies the identity \cite{Dubovsky2,BB}
\begin{equation}
\label{current_div} J^{\mu}_{~;\mu} \equiv 0.
\end{equation}
where semicolon indicates covariant differentiation.
The quantities $J^\mu $ and $b$ can be identified with the entropy
current and the entropy density correspondingly \cite{Dubovsky2}.

\subsection{ Higher derivative terms.}

The theory outlined above has been purely classical. The classical treatment of the effective field theory should be applicable on  scales of the order of $1/\Lambda$ or longer, where  $\Lambda$ is some positive constant. We assume that a cut-off parameter $ \Lambda $ is smaller than the Planck mass $ M_{PL}\equiv 1/\sqrt{8\pi G}$, but large enough to cover all relevant cosmological scales. To be able to neglect quantum fluctuations, the scalar fields $\varphi^a$ would be sufficiently smooth \cite{Dubovsky1}:
\begin{equation}
|\partial_\nu \varphi^a|\gg \frac{\partial_\mu \partial_\nu\varphi^a}{\Lambda}.
\end{equation}

This inequality allows us to use a derivative expansion and write the action of the theory in the form \cite{Dubovsky1, NLO}
\begin{equation}
S=\int \Lambda^4\tilde{F} \left(\epsilon\frac{\varphi^a}{\Lambda},\frac{B^{ab}}{\Lambda^4},  \frac{\varphi^{a;\mu}_{~~\mu}\varphi^{b;\nu}_{~~\nu}}{\Lambda^6}, ...\right)\sqrt{-g}d^4x,
\end{equation}
where  $\epsilon \ll 1$ is a small dimensionless parameter. It is assumed usually that all dimensionless parameters (except  $\epsilon$) in the function $\tilde{F}$ are of order one. This last assumption is very strong in some cases, we are not imposing it here. Nevertheless, we expect that a derivative expansion of $\tilde{F}$ is uniformly convergent at sufficiently smooth $\varphi^a$. In what follows we will work in units where $\Lambda =c = \hbar =1$.

Consider now possible corrections to the Lagrangian of a
barotropic perfect fluid. We will constraint the great variety of
possible additional terms by the assumption that our field theory
will contain no derivatives of scalar fields $\varphi^a$ higher
than second order. Within the framework of fluid description, such
terms can be constructed from hydrodynamic variables $ \rho $, $
u^\mu $ and their first order derivatives. Following
\cite{Dubovsky2}, we will group the corrections according to the
total number of involved derivatives of the hydrodynamic
variables.

The most general first-order term in the Lagrangian which is
compatible with the symmetries (\ref{symmetry1})-(\ref{symmetry3})
has the form
\begin{equation}
\label{f-o}f(b) b_{,\mu}u^{\mu},
\end{equation}
where $f(b)$ is an arbitrary function. It is worth noting that
this term is linear in the second derivatives of fields
$\varphi^a$. Since the equation (\ref{current_div})  is an
identity, the theory is free of first-order corrections
\cite{Dubovsky2,BB}. Indeed, in view of the equations
(\ref{current}) and (\ref{current_div}), the term (\ref{f-o}) can
be reduced to a total derivative.

The general form of second-order correction is a linear
combination of six scalars  \cite{Nondissipative,NLO}
\begin{equation}
\label{hd_all}{\begin{array}{*{20}c}
h_1(b)\left(u^{\mu} b_{,\mu}\right)^2, ~~~h_2(b) b^{,\mu} b_{,\mu},
~~~h_3(b) u^{\mu}_{~;\nu} u^{\nu}_{~;\mu},\\
h_4(b)\varepsilon_{\alpha\beta\mu\nu} u^{\alpha;\beta} u^{\mu;\nu}
,~~~ h_5(b) \left(u^{\mu}_{~;\mu}\right)^2, ~~~h_6(b) u^{\mu}
u^{\nu} u^{\alpha}_{~;\mu} u_{\alpha ;\nu},
\end{array}}
\end{equation}
where  $h_1(b),...,h_6(b)$ are arbitrary scalar functions, and
$\varepsilon_{\alpha\beta\mu\nu}$ is the alternating unit tensor.
Fluid Lagrangians involving the term $\gamma \left(u^{\mu}_{~;\mu}
\right)^2$ with constant $\gamma$ have been examined recently
\cite{MimeticHD1, MimeticHD2} in the context of mimetic gravity
\cite{MimeticGravity}. The more general models, in which $\gamma$
is a function of the mimetic field, are also investigated in
\cite{MimeticHD2}.

The case of second-order correction
\begin{equation}
\label{hd}f(b)b_{,\mu}b^{,\mu}
\end{equation}
appears to be the simplest to general study.  A
sample of such term with $f(b)= const$ has been considered in
\cite{Dubovsky2}. Moreover, the
term (\ref{hd}) plays a crucial role in the analysis of  dark matter perturbations  in the framework of the effective field theory for hydrodynamics (we will discuss this issue at the beginning of Section \ref{sec:4}).

The second-order term (\ref{hd}) in the Lagrangian gives the
following contribution to the energy-momentum tensor
\begin{equation}
\label{Thd}T_{\mu\nu (hd)}=-2f b_{,\mu}b_{,\nu}+ fb_{,\sigma}
b^{,\sigma}g_{\mu\nu} + \left(2f bb^{;\sigma} _{~~;\sigma}
+f_{,b}b^{,\sigma}b_{;\sigma} b\right) \left(g_{\mu\nu} + u_\mu
u_\nu\right).
\end{equation}
The variation of the total action with respect to scalar fields
$\varphi^a$  yields the equations of motion of the fluid
\begin{equation}
\label{fluid_base} \left[ \sqrt{|g|} g^{\mu\nu} b\left(F_{,b} -
f_{,b}b_{;\sigma} b^{;\sigma} -2f b^{;\sigma}_{~;\sigma}\right)
(B^{-1})_{ab} \varphi^a_{,\mu}\right]_{,\nu} =0.
\end{equation}

\subsection{ Model.}

In the following, we will study the case of the power-law
dependence of  $f(b)$.  The corresponding term in the Lagrangian
takes the form
\begin{equation}
\label{hdmodel} \alpha  b^n b_{,\mu}b^{,\mu}.
\end{equation}
This choice of second-order corrections  looks natural since the
description of any perfect fluid with constant equation of state
also uses the power-law dependence of the function $F(b)$.

To completely specify a model, one need to choose the lowest order
Lagrangian. In this paper, we consider a model of cold dark matter
with higher derivative corrections performing a more general
analysis where it is convenient or necessary. The cold dark matter
action with considered higher derivative corrections has the form
\begin{equation}
\label{action_cdm} S=\int\left(\alpha b^n b_{,\mu}b^{,\mu} + \beta
b\right) \sqrt{|g|}d^4x ,
\end{equation}
where $\alpha$ and $\beta$  are some constants. Instead the
coupling constant $ \alpha $, one can  use the dimensionless
parameter $8 \pi G\alpha$.

\section{Background.}
\label{sec:3}

In this Section we consider a homogeneous, isotropic and spatially
flat Universe. The line element has the form
\begin{equation}
\label{spf}ds^2=a^2\left(-d\tau^2+\delta_{ij}x^ix^j\right),
\end{equation}
where $a$ is the scale factor and $\tau$ is conformal time.

Since the fields $\varphi^a$ can be regarded as comoving
coordinates of the fluid, one can write
\begin{equation}
\bar{\varphi}^a=  x^a.
\end{equation}
Here, the bar indicates that the expression is evaluated for the
background line element (\ref{spf}).

Substituting the background fields $\bar{\varphi}^a$ into the
equation (\ref{matrix}), we obtain \cite{BB}
\begin{equation}
\label{barb}\bar{B}_{ab}=\frac{1}{a^2}\delta_{ab}, ~~~~~
\bar{b} =\frac{1}{a^3}.
\end{equation}
A direct check shows that the background fields $\bar{\varphi}^a$
are the solutions of the field equations (\ref{fluid_base}).

To study the background evolution of the fluid with higher
derivative corrections,  we use a slightly different approach than
that was implemented in \cite{NLO}. We split a full
energy-momentum tensor into two parts:
\begin{equation}
\label{decomp}T_{\mu\nu} =T_{\mu\nu (m)} +T_{\mu\nu (hd)}  ,
\end{equation}
where $T_{\mu\nu (hd)}$ is defined by the equation (\ref{Thd}).
When, the equations (\ref{decomp}) and (\ref{Thd}) give the
following expressions for the background energy-momentum tensor
components
\begin{eqnarray}
\label{T_backgr00}\bar{T}^{0}_{~0} &=& - \bar{\rho} +
9\alpha\frac{\mathcal{H}^2}{a^2}  a^{-3(n+2)},\\
\label{T_backgrij}\bar{T}^{i}_{~j} &=&  \left[ \bar{p}+  3\alpha
\frac{\mathcal{H}^2}{a^2} a^{-3(n+2)} \left(
2\frac{\dot{\mathcal{H}}}{\mathcal{H}^2} -3n-5 \right) \right]
\delta^{i}_{~j},
\end{eqnarray}
where $\mathcal{H}=\dot{a}/a$, the dot denote the derivative
respect to conformal time $\tau$, and
\begin{eqnarray}
\bar{\rho} &=& -\bar{T}^{0}_{~0 (m)},\\
\bar{p} &=& \bar{T}^{i}_{~j (m)}  ~~~(i=j).
\end{eqnarray}
The equations (\ref{T_backgr00}), (\ref{T_backgrij}) indicate that
the background energy-momentum tensor  depends on the quantity
$\mathcal{H}$ and its derivative with respect to the conformal
time. The energy-momentum  tensors with the same feature have been
studied  in cosmological models with a viscous fluid (recent
review can be found in \cite{bulk}).

The dynamics of the scale factor determined by the background
Einstein equations
\begin{equation}
\label{einst} \bar{G}^{\mu}_{~\nu} =8\pi G \bar{T}^{\mu}_{~\nu}  ,
\end{equation}
where $ \bar{G}^{\mu}_{~\nu}$ is the  background Einstein tensor
and $G$ is the gravitational constant. The non-zero components of
the Einstein tensor for spatially flat metric are
\begin{equation}
\bar{G}^{0}_{~0} = -3\frac{\mathcal{H}^{2}}{a^2}, ~~~~~
\bar{G}^{1}_{~1} = \bar{G}^{2}_{~2} =\bar{G}^{3}_{~3} =
-\frac{2\dot{\mathcal{H}} + \mathcal{H}^{2}}{a^2} .
\end{equation}

It is convenient to introduce the notation
\begin{equation}
\tilde{G}(a) = \frac{G}{1+24\pi G \alpha a^{-3(n+2)}} .
\end{equation}
In general, the quantity $\tilde{G}$ is a function of the scale
factor $a$. This function is constant at  $n=-2$.

Now the Einstein equations lead to
\begin{eqnarray}
\label{backgr1}\mathcal{H}^{2} &=& \frac{8\pi \tilde{G}}{3}
a^2\bar{\rho} ,\\
\label{backgr2}\dot{\mathcal{H}} &=& -\frac{4\pi \tilde{G}}{3} a^2
\bar{\rho}\left[\left(1 + 3 w\right) - \frac{72\pi \tilde{G}
\alpha}{a^{3(n+2)}}(n+2) \right].
\end{eqnarray}

The equation (\ref{backgr1}) allows  to interpret the quantity
$\tilde{G}$ as the effective background gravitational "constant".
In this interpretation, any theory with the term (\ref{hdmodel})
can be considered as a modified theory of gravity. At $ n = -2 $,
the equations  (\ref{backgr1}), (\ref{backgr2}) are reduced to the
corresponding Friedmann equations with a renormalized
gravitational constant.

For negative $\alpha$,  the quantities $\tilde{G}$ and
$\mathcal{H}$ can be infinite. It occurs when the scale factor
takes the critical value $a_s$, which is the solution of the
equation
\begin{equation}
\frac{24\pi G\alpha}{ a_s^{3(n+2)}}=-1 .
\end{equation}
The critical value $a_s$ is determined by the coupling constant
$\alpha$ and the power-law index $n$. At $n>-2$ and sufficiently
small $|\alpha|$, the singularity falls on the Planck era near
which our effective field theory  is not applicable. This singularity 
also can be avoided in more complex models that involve a generating
process of the dark matter fluid. At $n<-2$, the  role of term
(\ref{hdmodel}) increases with time and the singularity can be
eliminated in models with a decaying fluid.

For positive $\alpha$, it is possible to have
\begin{equation}
\frac{24\pi G\alpha}{ a^{3(n+2)}} \gg 1 .
\end{equation}
This regime realized for sufficiently small values of the scale
factor when  $n>-2$, or for sufficiently large values of the scale
factor when  $n<-2$. In this mode, we obtain $\tilde{G}/ G \approx
a^{3(n+2)}/(24\pi G\alpha)$, and the equation (\ref{backgr1}) is
significantly different from the usual second Friedman equation.
However, the near-standard evolution of the scale factor $a(\tau)$
during the radiation dominated and matter dominated stages can be
obtained only  when the influence of higher derivative terms on
the background solution is weak or veiled.

Hence, the treatment of only one type (\ref{hdmodel}) of
corrections  is applicable if we have $n\approx-2$ with a good
accuracy, or if the background effect of these corrections is
subdominant during considered time period
\begin{equation}
\label{subb}\frac{24\pi G\left|\alpha\right|}{ a^{3(n+2)}} \ll 1  .
\end{equation}

\section{Perturbations.}
\label{sec:4}

To begin with, let us consider the perturbations of the fields 
$\varphi^a$ in the Minkowski space-time. We use here the 
parametrization \cite{Dubovsky1, BB, NLO}
\begin{equation}
\varphi^a=x^a+\pi^a .
\end{equation}

By expanding the equations (\ref{matrix}), (\ref{def_b}), 
(\ref{velocity})  up to second order in $\pi$, one can 
obtain \cite{NLO}
\begin{equation}
b = 1+\partial_i\pi^i  -\frac{1}{2}\dot{\pi}^2-\frac{1}{2} 
\partial_i\pi^j\partial_j\pi^i +\frac{1}{2}(\partial_i\pi^i)^2
\end{equation}
and
\begin{equation}
u^0=1+\frac{1}{2}\dot{\pi}^2,~~~~~u^i=-\dot{\pi}^i + 
\dot{\pi}^k \partial_k\pi^i.
\end{equation}

For simplicity, we assume that the wavelengths of the 
perturbations are long enough and $\bar{\rho}\neq 0$. Then the 
quadratic action for dark matter perturbations, including 
\textit{all} higher derivative terms (\ref{hd_all}), takes the form
\begin{equation}
\label{quad_action}S^{(2)} = S^{(2)}_{LO} + S^{(2)}_{NLO},
\end{equation}
where \footnote{The equation (\ref{quad_action_NLO}) is exactly the 
same as the Eq. (3.20) of \cite{NLO} for considered here case  
$c_\pi^2\equiv \bar{p}'/\bar{\rho}' =0$, $\bar{p} =0$.}
\begin{eqnarray}
\label{quad_action_LO}S^{(2)}_{LO} &=& \frac{\bar{\rho}}{2}  \int   
\left(\dot{\pi}^2_L+\dot{\pi}^2_T\right)  d^4x ,\\
\label{quad_action_NLO}S^{(2)}_{NLO} &=& h_2 \int 
\left(\partial_j \left( \partial_i\pi^i\right)\right)^2 d^4x,
\end{eqnarray}
and we split the fields $\pi^i$ into the longitudinal and 
transverse parts:
\begin{equation}
\pi^i=\pi_L^i+\pi_T^i,~~~~~\partial_i\pi^i_T =0,
~~~~~\epsilon_{ijk}\partial_j\pi^k_L =0.
\end{equation}

The variation of the quadratic action (\ref{quad_action})  with 
respect to  the perturbations $\pi^i$ gives the equations of motion
\begin{eqnarray}
\label{eq_transverse}\ddot{\pi}_T^i  =0,\\
\label{eq_longitudinal}\ddot{\pi}^i_L - 2\frac{h_2}{\bar{\rho}} 
\partial^2 \left(\partial_k \partial_i\pi^i_L\right) =0.
\end{eqnarray}

The latter equation is conveniently rewritten using a new 
 variable $s$ defined by  \cite{NLO}
\begin{equation}
\pi^i_{L} =\partial_i s.
\end{equation}
The equation (\ref{eq_longitudinal}) yields in Fourier space
\begin{equation}
\label{eq_s}\ddot{s} - 2\frac{h_2}{\bar{\rho}} k^4 s =0.
\end{equation}
Hence, longitudinal perturbations do not grow exponentially 
with time when
\begin{equation}
\label{longitudinal_constr}\frac{h_2}{\bar{\rho}} \leq 0.
\end{equation}
For the model (\ref{action_cdm}) it gives the constraint
\begin{equation}
\label{sign_constr}\alpha\leq 0.
\end{equation}

Under the condition (\ref{longitudinal_constr}), the equation 
(\ref{eq_s}) describe a simple harmonic oscillator with 
k-dependent mass term which stem from higher derivative corrections. 
Consequently, in the effective field theory approach for dark 
matter, we expect $ \dot{\pi} \sim  \sqrt{h_2/\bar{\rho}}k^2\pi $ 
and $ S^{(2)}_{NLO} \sim S^{(2)}_{LO} $ due the equations of motion.

\subsection{Perturbed Einstein equations.}
\label{sec:4.1}

Let us consider small linear perturbations around a spatially flat
Friedmann-Robertson-Walker Universe, focusing on scalar
perturbations. It is convenient to use the freedom of gauge
transformations and fix the longitudinal gauge. The line element
in the longitudinal gauge has the form \cite{MFB}
\begin{equation}
ds^2 = a^2(\tau)\left\{-(1+2\Phi) d\tau^2 +(1 - 2\Psi) \delta
_{ij} dx^idx^j \right\},
\end{equation}
where $\Phi$ and $\Psi$ are some small scalar quantities which are
functions of space and time coordinates.

The scalar fields $\varphi^a$ also can be split into a background
part and a small perturbation
\begin{equation}
\varphi^a(\tau,x^i)=\bar{\varphi}^a(\tau,x^i)+ \delta \varphi^a
(\tau,x^i).
\end{equation}
Since $\bar{\varphi}^i(\tau,x^i)=x^i$, the equations
(\ref{matrix}), (\ref{def_b}) yield  \cite{BB}
\begin{equation}
\label{b_approx}b = \frac{1}{a^3} \left(1 + 3\Psi  +  \delta
\varphi ^k_{~,k} \right).
\end{equation}
The equation (\ref{velocity}) gives
\begin{equation}
\label{velocityscal} u^0 =  \frac{1}{a}\left(1-\Phi\right), ~~~~~
u^i = - \frac{1}{a} \delta\dot{\varphi}^i.
\end{equation}

We consider only the scalar perturbations. This means that, in
particular, one can introduce a velocity potential $v$ such that
$u^i=v^{,i}/a$. The equation (\ref{velocityscal}) then leads to
\begin{equation}
\delta\varphi^i  =\delta\chi^{,i},~~~~~ v  = -\delta\dot{\chi}.
\end{equation}
where $\delta\chi$ is some small scalar quantity.

The linearization of the energy-momentum tensor of relativistic
fluid with second-order corrections yields
\begin{eqnarray}
\label{thd00}\delta T^0_{~0}&=&-\delta \rho + 6f\mathcal{H}^2
\frac{\bar{b}^2}{a^2} \left[  3  \frac{\delta b}{\bar{b}} -3\Phi -
\frac{1}{\mathcal{H}} \left(  \frac{\delta b}{\bar{b}} \right)
^{\bcdot}+\frac{3}{2}\frac{f_{,b}}{f} \delta b\right], \\
\label{thd0i}\delta T^0_{~i} &=&  \left( \bar{\rho} + \bar{p}
\right) v - 6f\mathcal{H}^2\frac{\bar{b}^2}{a^2} \left[
\frac{1}{\mathcal{H}} \left(  \frac{\delta b}{\bar{b}}\right)_{,i}
-\left( 1- \frac{\dot{\mathcal{H}}}{\mathcal{H}^2} + \frac{3}{2}
\frac{f_{,b}}{f}\bar{b} \right) \delta\dot{\varphi}_i\right],\\
\label{thdij}\delta T^i_{~j} &=& \delta \bar{p} \delta^{i}_{~j} +
2f\frac{\bar{b}^2}{a^2}  \Bigg[- \left(  \frac{\delta b}{\bar{b}}
\right)^{\bcdot\bcdot} +\mathcal{H} \left( 7+3\frac{f_{,b}}{f}
\bar{b} \right)  \left( \frac{\delta b}{\bar{b}}\right) ^{\bcdot}
\nonumber \\
&&+~~ \mathcal{H}^2 \left(3\left(2 \frac{\dot{\mathcal{H}}
}{\mathcal{H} ^2} -5 \right) +3\frac{f_{,b}}{f}\bar{b} \left(
\frac{\dot{\mathcal{H}}}{\mathcal{H}^2}-7 \right) -\frac{9}{2}
\frac{f_{,bb}}{f}\bar{b}^2\right)\frac{\delta b}{\bar{b}}
\nonumber \\
&& +~~ \nabla^2\frac{\delta b}{\bar{b}}+3\mathcal{H}^2 \left( 5 -
2 \frac{\dot{\mathcal{H}}}{\mathcal{H}^2} + 3\frac{f_{,b}}{f}
\bar{b} \right) \Phi - 3\mathcal{H} \left(3\dot{\Psi} +\dot{\Phi}
\right) \Bigg]\delta ^i_{~j},
\end{eqnarray}
where we denote
\begin{equation}
\delta\rho = -\bar{F}_{,b}\delta b ,~~~~~ \delta p = -
\bar{b}\bar{F}_{,bb} \delta b .
\end{equation}

Using equations  (\ref{b_approx}) and (\ref{thd00})-(\ref{thdij}),
one can write the perturbed Einstein equations \cite{MFB}.  For
the case $f=\alpha b^n$, they are
\begin{eqnarray}
&&3\mathcal{H}(\dot{\Phi} + \mathcal{H}\Phi) -\nabla^2\Phi =
-4\pi G a^2 \delta \rho\nonumber \\
\label{einst1} && \qquad ~~ - ~ \frac{24\pi G\alpha}{ a^{3(n+2)}}
\left[ 3\mathcal{H}(\dot{\Phi} + \mathcal{H}\Phi)  +  \mathcal{H}
\nabla^2  \delta \dot{\chi} - \frac{3}{2}\mathcal{H}^2\left(n
+2\right)\left( 3\Phi  + \nabla^2  \delta \chi\right)\right],\\
&&\dot{\Phi} + \mathcal{H}\Phi = -4\pi G a^2(\bar{\rho} + \bar{p})
v \nonumber \\
\label{einst2} &&\qquad \qquad\qquad\qquad \!\!\!\! - ~ \frac{24
\pi G\alpha}{a^{3(n+2)}}  \left[ \frac{1}{2}\left( 3 n + 2- 2
\frac{\dot{\mathcal{H}}}{\mathcal{H}^2} \right) \mathcal{H}^2
\delta\dot{\chi} -\mathcal{H} \left( 3\Phi   + \nabla^2  \delta
\chi\right) \right],\\
&&\ddot{\Phi} + 3\mathcal{H}\dot{\Phi} + \left(
{2\dot{\mathcal{H}} + \mathcal{H}^{2}} \right)\Phi
=4\pi G\delta p \nonumber \\
&&~~ - ~ \frac{24\pi G\alpha}{a^{3(n+2)}} \left[3\ddot{\Phi}\!-9\!
\left(1+n\right)\mathcal{H}\dot{\Phi}\!+\! \left(6 +\frac{9}{2}
n\right) \!\left(3n+\!5\! -2\frac{\dot{\mathcal{H}}}{
\mathcal{H}^2} \right) \mathcal{H}^2 \Phi \! - \! 3\nabla^2 \Phi
\right. \nonumber \\
\label{einst3} &&~~ +  \left. \nabla^2 \! \left(\delta \ddot{\chi}
- \left(7+3n\right) \mathcal{H}    \delta \dot{\chi} + \frac{3}{2}
(n+2) \left(3n+\! 5\! -2\frac{\dot{\mathcal{H}}}{\mathcal{H}^2}
\right) \mathcal{H}^2 \delta \chi  -  \nabla^2  \delta \chi\right)
\right]\!.
\end{eqnarray}

Here we used the equality $\Psi = \Phi$ which follows from the
fact that the spatial part of the energy-momentum tensor is
diagonal. In this form,  the perturbed Einstein equations can be
easily generalized to the case of multiple minimally coupled
fluids with vanishing anisotropic stresses.

The higher derivative terms may cause singularities in the
solution of the perturbed equations.  These singularities can be
avoided by tuning the  model parameters or initial conditions. For
the rest of this section  we investigate the constraints on the
parameters, which follow from the condition of the absence of
instabilities during the radiation dominated and matter dominated
stages.

\subsection{Instabilities in the dust dominated Universe.}
\label{sec:4.2}

Consider a Universe filled with ordinary non-relativistic matter
$(b)$ and cold dark matter with higher derivative corrections
$(c)$. For our purposes, the most convenient equations  are the
Einstein equations (\ref{einst1}),  (\ref{einst2}), in which  we
have to make the substitution
\begin{eqnarray}
\delta\rho &=&\delta \rho_b -\beta\delta b,\\
\label{vdust}\left(\bar{\rho}+\bar{p}\right)v &=& \bar{\rho}_b v_b
+ \beta \bar{b}\delta\dot{\chi}.
\end{eqnarray}

Since
\begin{equation}
\label{rho_c} \bar{\rho}_c \equiv -\beta\bar{b},
\end{equation}
the equation (\ref{vdust}) takes the form of the well-known
expression for the case of multi-component fluid
\begin{equation}
\left(\bar{\rho}+\bar{p}\right)v = \sum_A\left(\bar{\rho}_A +
\bar{p}_A\right)v_A,
\end{equation}
where subscript  $A$ denotes the fluid species.

The equations (\ref{barb}), (\ref{backgr1}), (\ref{rho_c}) lead to
\begin{equation}
4\pi G \frac{\beta}{a} =-\frac{3}{2}\frac{G}{\tilde{G}}
\frac{\bar{\rho}_c}{\bar{\rho}} \mathcal{H}^2.
\end{equation}

In addition, there is the equality
\begin{equation}
\bar{\rho} =\bar{\rho}_b+\bar{\rho}_c .
\end{equation}

The independent perturbed Einstein equations  can be written now
in the Fourier space as
\begin{eqnarray}
&& \dot{\Phi}  -\frac{8\pi \tilde{G}\alpha}{ a^{3(n+2)}}  k^2
\delta \dot{\chi} =-\frac{1}{2}\frac{\rho_b}{\bar{\rho}}
\mathcal{H} \delta_b   + \left(\frac{1}{2} \frac{\bar{\rho}_c
}{\bar{\rho}} - \frac{12\pi \tilde{G}\alpha}{ a^{3(n+2)}}  \left(n
+2\right)\right)\mathcal{H} k^2  \delta \chi \nonumber \\
\label{einstcdm1} &&\qquad\qquad\qquad\qquad  - \left( 1 +
\frac{3}{2} \frac{\bar{\rho}_c}{\bar{\rho}}  + \frac{k^2}{3
\mathcal{H}^2} \frac{\tilde{G}}{G} - \frac{36\pi \tilde{G}\alpha}{
a^{3(n+2)}}  \left(n +2\right)\right) \mathcal{H}\Phi, \\
&& \dot{\Phi} -\left( \frac{3}{2}\frac{G}{\tilde{G}} \frac{
\bar{\rho} _c}{\bar{\rho}}-\frac{12\pi G\alpha}{a^{3(n+2)}} \left(
3 n + 2- 2\frac{\dot{\mathcal{H}}}{\mathcal{H}^2} \right)\right)
\mathcal{H}^2 \delta\dot{\chi} = \nonumber \\
\label{einstcdm2} &&\qquad \qquad ~~~ -\frac{3}{2}
\frac{G}{\tilde{G}} \frac{\bar{\rho}_b}{\bar{\rho}}\mathcal{H} ^2
v_b - \frac{24\pi G\alpha}{a^{3(n+2)}} \mathcal{H} k^2  \delta
\chi - \left(1  - \frac{72\pi G\alpha}{a^{3(n+2)}} \right)
\mathcal{H}\Phi    .
\end{eqnarray}

These equations can be solved with respect to $\delta\dot{\chi}$ 
and $\dot{\Phi}$. For example, subtracting  (\ref{einstcdm2}) 
from (\ref{einstcdm1}), we obtain
\begin{eqnarray}
&& \left(
\frac{3}{2}\frac{G}{\tilde{G}}\frac{\bar{\rho}_c}{\bar{\rho}}
-\frac{12\pi G\alpha}{a^{3(n+2)}} \left( 3 n + 2-
2\frac{\dot{\mathcal{H}}}{\mathcal{H}^2} \right) - \frac{8\pi
\tilde{G}\alpha}{ a^{3(n+2)}}  \frac{k^2 }{\mathcal{H}^2} \right)
\mathcal{H} \delta\dot{\chi}= \frac{3}{2}\frac{G}{\tilde{G}}
\frac{\bar{\rho}_b}{\bar{\rho}}\mathcal{H} v_b \nonumber \\
&& \qquad \!\!\! - \frac{1}{2}\frac{\rho_b}{\bar{\rho}}\delta_b
-\frac{k^2}{3\mathcal{H}^2} \frac{\tilde{G}}{G}\Phi  -\!\left(\!
\frac{1}{2}\frac{\bar{\rho}_c}{\bar{\rho}}+ \frac{24\pi
G\alpha}{a^{3(n+2)}} -\frac{12\pi \tilde{G}\alpha}{ a^{3(n+2)}}
\left(n +2\right)\! \right)\left( 3\Phi  - k^2  \delta
\label{singular_eq}\chi\right)\!.
\end{eqnarray}

The equation (\ref{singular_eq}) indicates that the conformal 
time derivative of the quantity $\delta\chi$  is ill-defined 
at the moment $\tau_s$ at which
\begin{equation}
\label{singular} \frac{3}{2}\frac{G}{\tilde{G}} \frac{\bar{\rho}
_c}{\bar{\rho}} -\frac{12\pi G\alpha}{a^{3(n+2)}} \left( 3 n + 2-
2\frac{\dot{\mathcal{H}}}{\mathcal{H}^2} \right) - \frac{8\pi
\tilde{G} \alpha}{ a^{3(n+2)}}  \frac{k^2 }{\mathcal{H}^2}    =0.
\end{equation}

Provided that the quantities  $\delta\dot{\chi}$ and $\dot{\Phi}$ 
are finite, the equation (\ref{singular_eq}) determines 
$\Phi(\tau_s)$ as a function of $v_b(\tau_s)$, $\delta_b(\tau_s)$, 
$\delta \chi(\tau_s)$. Under the same assumption, taking the 
conformal time derivative of (\ref{einstcdm1}), (\ref{einstcdm2}) 
and performing some algebraic calculations, we can uniquely express 
$\delta\dot{\chi}(\tau_s)$, $\dot{\Phi}(\tau_s)$ in terms of these 
three quantities. Thus, the non-singularity of the solution at the 
moment $\tau_s$ requires extremely fine tuning of  initial conditions.

The background equations yield $\mathcal{H}\approx 2/\tau$ in the
dust dominated Universe   (this equation becomes exact for $n=-2$
or $\alpha =0$). Under this approximation, the equation
(\ref{singular}) gives
\begin{equation}
\label{point_dust}k^2 \tau^2 =\frac{3}{4}\frac{G}{\tilde{G}}\frac{
a^{3(n+2)}}{\pi \tilde{G}\alpha}\frac{\bar{\rho}_c}{\bar{\rho}} -
18 \frac{G}{\tilde{G}} \left(  n + 1 \right) .
\end{equation}
The equation (\ref{point_dust}) has no solution at any $k$ if and
only if its right-hand side is negative. In this case the solution
of the complete system of perturbed equations is free of
singularities caused by ill-defined quantities $\delta\dot{\chi}$ 
and $\dot{\Phi}$.

At $\alpha\leq 0$, $n\geq -1$, the right-hand side of the equation
(\ref{point_dust}) is negative and it has no solution.  At
$\alpha\leq 0$, $n< -1$, the condition of the absence of
singularities takes the form
\begin{equation}
\frac{24\pi\tilde{G} \left|\alpha \right|}{ a^{3(n+2)}}
\leq\frac{1}{| n + 1 |}\frac{\bar{\rho}_c}{\bar{\rho}_b
+\bar{\rho}_c} .
\end{equation}

For positive $\alpha$, the right side of the equation
(\ref{point_dust}) can be negative only at $n > -1$  as long as
\begin{equation}
\label{constr2} \frac{24\pi\tilde{G}\alpha}{ a^{3(n+2)}} >
\frac{1}{  n +1 }\frac{\bar{\rho}_c}{\bar{\rho}_b+\bar{\rho}_c}.
\end{equation}

\subsection{Instabilities in the radiation dominated Universe.}
\label{sec:4.3}

Given that the Universe contain dark matter and radiation, the
analysis is very similar to the studied above case of dust
dominance. Now, the total density perturbation $\delta \rho$ and
the velocity potential $v$ are given by
\begin{eqnarray}
\delta\rho &=&\delta \rho_r -\beta\delta b, \\
\label{vrad} \left(\bar{\rho}+\bar{p}\right)v &=& \frac{4}{3}
\bar{\rho}_r v_r +\beta \bar{b}\delta\dot{\chi}.
\end{eqnarray}
The position of the singular point of the system of perturbed
equations is determined formally by the same equation
(\ref{singular}).

Hereinafter we  consider that the contribution of cold dark matter
to the total energy density is small, i.e. $\bar{\rho}_c \ll
\bar{\rho}$. We assume also that the  higher derivative terms do
not break the approximate equality $\mathcal{H}\approx 1/\tau$,
which should be fulfilled at the radiation dominated stage. This
allows us to rewrite the equation (\ref{singular}) as
\begin{equation}
\label{point_rad}k^2\tau^2   =\frac{3}{16}\frac{ a^{3(n+2)}}{\pi
\tilde{G} \alpha}\frac{G}{\tilde{G}} \frac{\bar{\rho}_c}{
\bar{\rho}}- \frac{3}{2}\frac{G}{\tilde{G}}\left( 3 n + 4 \right).
\end{equation}

At $\alpha \leq 0$,  $n \geq -\frac{4}{3}$ , the equation
(\ref{point_rad})  has no solution. At $\alpha \leq 0$,
$n<-\frac{4}{3}$, the solution does not exist under the condition
\begin{equation}
\label{subp}\frac{24\pi \tilde{G}\left|\alpha \right|}{
a^{3(n+2)}} \leq\frac{1}{\left|  n +\frac{4}{3} \right|}
\frac{\bar{\rho}_c}{\bar{\rho}_r + \bar{\rho}_c}.
\end{equation}
In the particular case $n=-2$, the equation (\ref{subp}) leads to
the equation (\ref{subb}) due the inequality $\bar{\rho}_c/
\bar{\rho}_r \ll 1$.

For positive $\alpha$, the equation (\ref{point_rad}) has a
solution only when $n > -\frac{4}{3}$ and there is the  additional
constraint
\begin{equation}
\label{constr1}\frac{24\pi\tilde{G}\alpha}{ a^{3(n+2)}} >\frac{1}{
n +\frac{4}{3} }\frac{\bar{\rho}_c}{\bar{\rho}_r+\bar{\rho}_c}.
\end{equation}
The equations (\ref{constr1}) and (\ref{constr2}), taken together,
are incompatible with the equation (\ref{subb}) which holds at
$n\neq -2$.

Now it is possible to summarize briefly some of our findings. The
condition of absence of singularities in both radiation dominated
and matter dominated epochs leads to the inequality (\ref{sign_constr}) 
together with the constraint (\ref{subb}) at any power-law index $n$.

\section{Evolution of perturbations.}
\label{sec:5}

Let us study the evolution of perturbations in a cosmological
fluid with higher derivative corrections in more detail. To
accomplish this, we consider two simple models. Using the results
of the previous section, one can restrict the consideration to the
case of non-positive coupling constant $\alpha$.

\subsection{Toy model.}

In this Toy model, the Universe contain only dark matter with the
action (\ref{action_cdm}). The evolution of perturbations is
governed by perturbed Einstein equations  (\ref{einst1}),
(\ref{einst2}). Using the background Einstein equations
(\ref{backgr1}) and (\ref{backgr2}), one can write them as
\begin{eqnarray}
&&\dot{\Phi}  -\frac{8\pi \tilde{G}\alpha}{ a^{3(n+2)}}  k^2
\delta \dot{\chi} =-\left(\frac{5}{2} +\frac{1}{3}
\frac{\tilde{G}}{G} \frac{k^2}{\mathcal{H}^2}-\frac{36\pi
\tilde{G}\alpha}{ a^{3(n+2)}}  \left(n +2\right)\right)
\mathcal{H} \Phi \nonumber \\
\label{toy1} &&\qquad \qquad\qquad\qquad \qquad\qquad
+\left(\frac{1}{2} -\frac{12\pi \tilde{G}\alpha}{ a^{3(n+2)}}
\left(n +2\right)\right)\mathcal{H} k^2  \delta \chi ,\\
&&\dot{\Phi} -\frac{3}{2}\left(1 - n\frac{24\pi \tilde{G}
\alpha}{a^{3(n+2)}}   + 2\frac{G}{\tilde{G}}\left(\frac{24\pi
\tilde{G}\alpha}{a^{3(n+2)}}\right)^2 \right)\mathcal{H}^2
\delta\dot{\chi} \nonumber \\
\label{toy2}&&\qquad \qquad\qquad = - \left(1-3\frac{G}{\tilde{G}}
\frac{24\pi \tilde{G}\alpha}{a^{3(n+2)}}  \right)\mathcal{H} \Phi
- \frac{G}{\tilde{G}}\frac{24\pi \tilde{G}\alpha}{a^{3(n+2)}}
\mathcal{H}  k^2  \delta \chi .
\end{eqnarray}

The equations (\ref{toy1}),(\ref{toy2}) can be combined into a
second order differential equation  for the metric perturbations
$\Phi$.  This equation has the form
\begin{equation}
\label{Phieq}f_1\ddot{\Phi }+ f_2\dot{\Phi } + f_3\Phi =0.
\end{equation}
The coefficients $f_1$, $f_3$, $f_3$ are some functions of
background quantities. Under condition (\ref{subb}), they are
\begin{eqnarray}
f_1 &=& 1-\frac{2}{9}\frac{k^2}{\mathcal{H}^2}\xi -\frac{8}{81}
\frac{k^4}{\mathcal{H}^4}\xi^3 ,\\
f_2 &=& 3\mathcal{H}\Bigg[1 + \frac{2}{27}\left(2+3n+(52+34n+3n^2)
\xi \right)\frac{k^2}{\mathcal{H}^2} \xi- \frac{8}{81} (3+n)
\frac{k^4}{\mathcal{H}^4} \xi^3\Bigg],\\
f_3 &=& -\mathcal{H}^2\Bigg[3  (n+2)^2 +\frac{1}{9} \left((3n+5)^2
+4 \right)\frac{k^2}{\mathcal{H}^2} +\frac{2}{9} \frac{k^4}{
\mathcal{H}^4} + \frac{8}{81} \frac{k^6}{\mathcal{H}^6} \xi^2
\Bigg] \xi,
\end{eqnarray}
where $\xi =24\pi G\alpha/a^{3(n+2)}$.

At $\alpha =0$, the equation (\ref{Phieq}) is reduced to
\begin{equation}
\label{dust_eq}\ddot{\Phi }+ \frac{6}{\tau}\dot{\Phi }  =0,
\end{equation}
The general solution of this equation  is
\begin{equation}
\label{dust_sol}\Phi_{dust} = C_1+\frac{C_2}{\tau^5},
\end{equation}
where $C_1, C_2$ are arbitrary constants. The constant solution
corresponds to the growing adiabatic mode at the  matter dominated
stage \cite{Durrer}
\begin{equation}
\label{dust_nondec}\Phi= C_1, ~~~~~\delta =-2 C_1 -\frac{1}{6} C_1
(k\tau)^2, ~~~~~v = -\frac{1}{3} C_1 k\tau ,
\end{equation}
where $\delta$ is the dust density contrast in the longitudinal
gauge.

At  $\alpha \neq 0$, the equation (\ref{Phieq}) can be solved in
several  special cases.

\subsubsection{Short-wavelength perturbations.}

Let us consider first the short-wavelength perturbations ($k\gg
\mathcal{H}$), that satisfy to the additional condition
\begin{equation}
\label{short2cond}\frac{24\pi G\left|\alpha\right|}{a^{3(n+2)}}
\frac{k^2}{\mathcal{H}^2}\ll 1.
\end{equation}
On this scale, the equation (\ref{Phieq}) is reduced to the form
\begin{equation}
\label{short2eq}\ddot{\Phi}+ 3\mathcal{H}\dot{\Phi } -
\frac{16}{3} \mathcal{H}^2 \frac{\pi G\alpha}{a^{3(n+2)}}
\frac{k^4}{\mathcal{H}^4}\Phi =0.
\end{equation}

To simplify the equation (\ref{short2eq}), one can use the
approximation $a\approx \tau^2/\tau_0^2$, $\mathcal{H} \approx
2/\tau$. Then the condition (\ref{short2cond}) can be rewritten as
\begin{equation}
\label{short2app}\frac{\pi G|\alpha
|\tau_0^{6(n+2)}}{\tau^{2(3n+5)}} k^2\ll 1.
\end{equation}
The  expression (\ref{short2app}) shows the importance of the
regime (\ref{short2cond}). At given $k$ and $n \geq -5/3$, if the
condition (\ref{short2app})  is fulfilled at some time, it must be
fulfilled at all subsequent times.

The characteristic time scale is equal to $\mathcal{H}^{-1}$.
Consequently, the mass term in the equation (\ref{short2eq}) can
be neglected when
\begin{equation}
\label{short2condstrong}\frac{24\pi G\left|\alpha \right|
}{a^{3(n+2)}} \frac{k^4}{\mathcal{H}^4}\ll 1.
\end{equation}
This fact is confirmed by the analysis of particular cases. The
inequality (\ref{short2condstrong}) itself, within the above
approximation, takes the form
\begin{equation}
\label{short2appstrong}\frac{\pi G|\alpha |\tau_0^{6(n+2)}}{
\tau^{2(3n+4)}} k^4\ll 1.
\end{equation}

For example, at $n=0$,  the equation (\ref{short2eq}) takes the
form
\begin{equation}
\ddot{\Phi } + \frac{6}{\tau}\dot{\Phi } -  \frac{4\pi G\alpha}{3}
\frac{k^4 \tau_0^{12}}{\tau^{10}}\Phi =0.
\end{equation}
The solution of this equation can be expressed in terms of Bessel
functions $J_{\nu}$ and $Y_{\nu}$ \cite{AbramowitzStegun}. The
general solution has the form
\begin{equation}
\Phi_{(0)} = \frac{C_{1(0)}}{\tau^{5/2}}J_{-5/8}\left(\sqrt{-
\frac{\pi G\alpha}{12} }\frac{\tau_0^{6}k^{2}}{\tau^4}\right) +
\frac{C_{2(0)}}{\tau^{5/2}}Y_{-5/8}\left(\sqrt{- \frac{\pi
G\alpha}{12} }\frac{\tau_0^{6}k^{2}}{\tau^4}\right),
\end{equation}
where $C_{1(0)}$ and $C_{2(0)}$ are some constants. This
expression oscillates and decreases rapidly  when
\begin{equation}
\frac{\pi G|\alpha |\tau_0^{12}k^4}{\tau^{8}} \gg 1 .
\end{equation}
Later, under the condition (\ref{short2appstrong}), the general
solution is not very different from the standard one
(\ref{dust_sol}), that can be checked by expanding the Bessel
functions into Taylor series.

Let us consider now the perturbations with extremely short
wavelength, such that
\begin{equation}
\label{short1cond}\frac{24\pi G\left|\alpha\right|}{a^{3(n+2)}}
\frac{k}{\mathcal{H}}\gg 1 .
\end{equation}
At any fixed moment of time, provided that $ \alpha \neq 0 $, one
can specify a small enough scale at which the condition above is
fulfilled. The equation (\ref{Phieq}) takes the form
\begin{equation}
\label{short1eq} \ddot{\Phi }+ 3(3+n)\mathcal{H}\dot{\Phi } +
k^2\Phi =0.
\end{equation}

The solution of the equation (\ref{short1eq}) can be found by the
WKB method. It has the form of oscillations which are damped at
$n>-3$ or increased at $n<-3$. Since fast and unlimited growth (in
the linear perturbation theory) of short-wavelength perturbations
destroys the spatial homogeneity of the Universe, we obtain the
constraint $n\geq-3$.

\subsubsection{Long-wavelength perturbations.}

On a large scale ($k\ll \mathcal{H}$), the equation (\ref{Phieq})
is reduced to
\begin{equation}
\ddot{\Phi} + 3\mathcal{H}\dot{\Phi } - \mathcal{H}^2\Bigg[3 (n+2)
^2 +\frac{1}{9} \left((3n+5)^2+4 \right) \frac{k^2}{\mathcal{H}^2}
\Bigg] \frac{24\pi G\alpha}{a^{3(n+2)}} \Phi =0.
\end{equation}

At  $n=-2$, the long-wavelength solutions of this equation have
the standard asymptotic behavior. At arbitrary  $n$, by virtue of
the equation (\ref{subb}), the mass term in this equation can be
neglected. Then we obtain, with reasonable accuracy,  the same
equation for the metric perturbations as in the absence of higher
derivative corrections.

The consideration of the Toy model shows that the characteristic
tendency of the model is the suppression of short-wavelength
perturbations. This feature, depending on the values of the
parameters $\alpha$ and $n$, can lead to a cutoff of the power
spectrum on cosmological scales. On the other hand, it may offer a
way to explain the uniform distribution of the dark matter within
the Solar system \cite{PP} by influence of higher derivative
terms.

\subsection{Matter dominated Universe.}

Compared with the Toy model, more realistic models would include
baryonic matter. In this subsection, we will study the evolution
of perturbations in the cosmological model, briefly considered in
Section \ref{sec:4.2}.

In  the longitudinal gauge, the Euler equation and the continuity
equation for baryonic matter are \cite{Durrer,Ma_Bertschinger}
\begin{eqnarray}
\label{barion1}\dot{\delta_b }  -3 \dot{\Phi} - k ^{2}v_b &=& 0,
\\
\label{barion2}\dot{v_b} + \mathcal{H}v_b - k^2 \Phi &=& 0.
\end{eqnarray}

The equations (\ref{barion1}), (\ref{barion2}) combined with the
perturbed Einstein equations (\ref{einstcdm1}),  (\ref{einstcdm2})
form a complete set of equations. This system can be solved
numerically by setting the initial adiabatic conditions on a large
scale. At $n \geq -4/3$, due to imposed initial conditions, the
condition (\ref{short2condstrong}) will be fulfilled during the
entire evolution and the impact of the corrections on the solution
of the system is negligible. At $ -4/3 < n < -3$  , the behaviour
of the solution is more complex. We numerically analyze it only
for the particular case $n=-2$.

\begin{figure}
\includegraphics[width=0.48\textwidth]{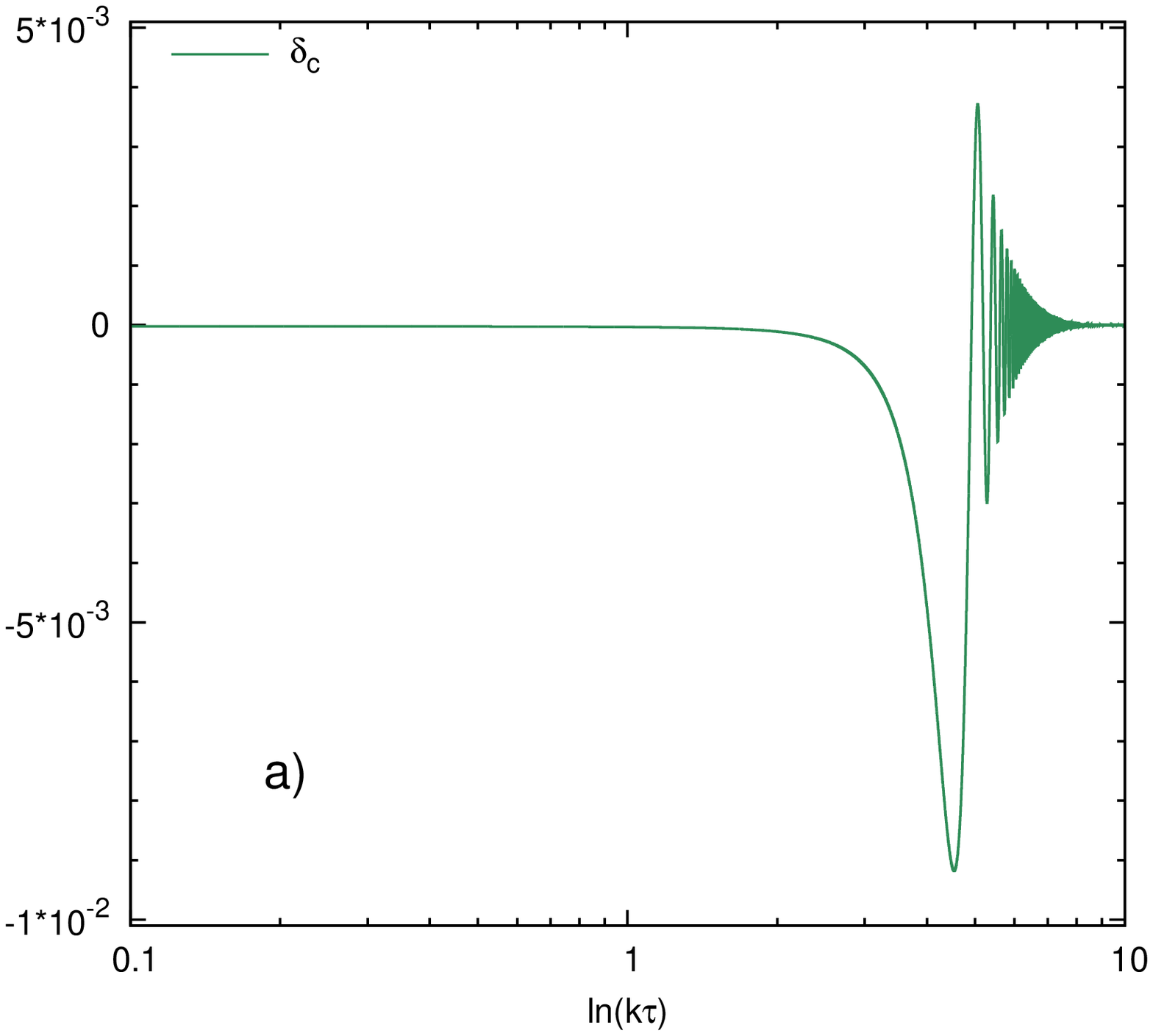}
\includegraphics[width=0.48\textwidth]{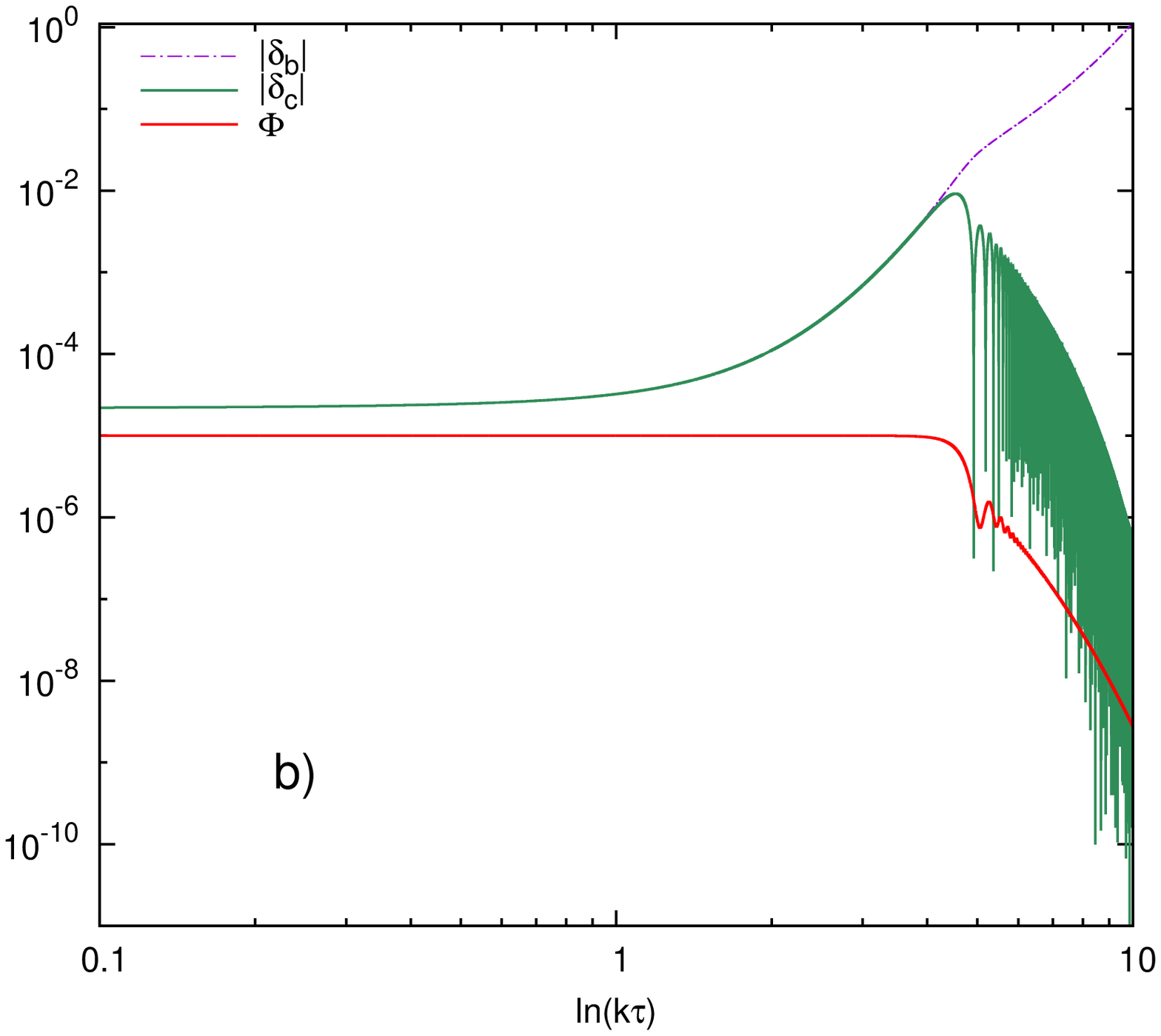}
\caption{a)The evolution of the quantity  $\delta_c \equiv
-k^2\delta \chi + 3\Phi$, which is, at $\alpha =0$, the  dark
matter density contrast in the longitudinal gauge. b) The
evolution of quantities  $|\delta_b |$, $|\delta_c |$, $\Phi$ on a
logarithmic plot. The solution corresponds to $8\pi G \alpha
=-10^{-6}$, $\frac{\bar{\rho}_c}{\bar{\rho}}=0.8$,
$\frac{\bar{\rho}_b}{\bar{\rho}}=0.2$. The initial conditions are
imposed  at $\ln (k \tau_{in}) = -40$. Plots are normalized by
$\Phi_0 =10^{-5}$. } \label{fig:1}
\end{figure}

It is convenient to introduce new variables  $x = k\tau$, $u_b=k
v_b  $, $D_c = -k^2\delta \chi$, $D_b=\delta_b- 3\Phi$.  The
governing equations when take the form
\begin{eqnarray}
D_b' &=& u_b ,\\
u_b' &=& - \frac{2}{x} u_b - \Phi,\\
\Phi'  + 8\pi \tilde{G}\alpha   D_c' &=& -\frac{1}{x} \left[
\frac{\bar{\rho}_c}{\bar{\rho}}D_c +\!\frac{\rho_b}{\rho}D_b
+\!\left(\!5 \!+\! \frac{1}{6} \frac{\tilde{G}}{G}x^2 \!\right)
\!\Phi \right] ,\\
\Phi' + \frac{6}{x^2}\left(\frac{\bar{\rho}_c}{\bar{\rho}}+24\pi
G\alpha \left( 1+\frac{\bar{\rho}_c}{\bar{\rho}}  \right)  \right)
D_c' &=&
-\frac{6}{x^2}\frac{G}{\tilde{G}}\frac{\bar{\rho}_b}{\rho}u_b
\nonumber \\
&&+ \frac{2}{x}\left[ \left(72\pi G\alpha  \!  -\! 1\right) \Phi
+ 24\pi G\alpha   D_c\right]\! .
\end{eqnarray}

We choose the usual adiabatic conditions,  which corresponds to
the growing adiabatic mode in case of zero corrections
(\ref{dust_nondec})
\begin{equation}
\Phi_{in}=\Phi_0, ~~~~~D_{b~\!in}=D_{c~\!in}=-5\Phi_0
-\frac{1}{6}\Phi_0 x_{in}^2, ~~~~~u_{b~\!in} = -\frac{1}{3}\Phi_0
x_{in} .
\end{equation}

The numerical solutions confirm the results of the above
analytical consideration. On a large scale, the impact of the
higher derivative terms is negligible. On a small scale, the dark
matter perturbations are suppressed passing through several
regimes. On a sufficiently small scale, the evolution of the
metric perturbations is governed by the density contrast of the
baryonic matter. The typical result (for $0<-8\pi G\alpha\ll 
\bar{\rho}_c/\bar{\rho}$) is shown in Figure \ref{fig:1}.

\section{Conclusions and outlook.}
\label{sec:6}

We examined some simplest possible higher derivative corrections
to the action of relativistic fluid within the  effective field
theory for hydrodynamics. Considered corrections can be described
by two parameters, namely the coupling constant $\alpha$ and the
power-law index $n$. Particular attention has been given to dark
matter, described by the fluid action  (\ref{action_cdm}). The
condition of absence of singularities in the background and
perturbed solutions  leads to the constraint $\alpha\leq 0$, as
well as to the inequality  $8\pi G\alpha/a^{3(n+2)}\ll 1$ that
should be fulfilled during the radiation dominated and the matter
dominated stages. The evolution of perturbations has been
investigated in two models of the matter dominated Universe. We
obtain what considered higher derivative corrections have a strong
influence on the evolution of  short-wavelength perturbations.
Consideration of the Toy model (without baryonic matter) shows
that the perturbations are suppressed in the short-wavelength
limit at $ n> -3 $. The numerical analysis in a more realistic
model with baryonic matter confirms and refines results of the
analytical treatment. In this model, there is a wide range of
parameters in which the dark matter can be considered as
homogeneous on a sufficiently small scale.


\subsection*{Acknowledgements}

\bigskip
The author is grateful to  Alexander Vikman for drawing attention
to the references \cite{MimeticHD1, MimeticHD2}.
\bigskip\

\end{document}